\documentclass[sigconf]{acmart}
\AtBeginDocument{%
  }

\setcopyright{acmlicensed}
\copyrightyear{2025}
\acmYear{2025}
\setcopyright{acmlicensed}
\acmConference[Anonymous]{Anonymous Conference}{Month Day--Day, 2025}{Location}
\acmBooktitle{Proceedings of the Anonymous Conference, Month Day--Day, 2025, Location}

\usepackage{tabularx}
\usepackage{caption}
\usepackage{subcaption}
\usepackage{booktabs}
\usepackage{soul}
\usepackage{multirow}
\usepackage{makecell}
\usepackage{balance}
\usepackage{float}
\usepackage{longtable}
\usepackage{lipsum}
\usepackage{setspace}
\usepackage{subcaption}

\begin{document}

\title[Auditing Political Exposure Bias on Twitter/$\mathbb{X}$]{Auditing Political Exposure Bias: Algorithmic Amplification on Twitter/$\mathbb{X}$ During the 2024 U.S. Presidential Election}


\author{Jinyi Ye}
\orcid{0009-0004-7757-1642}
\affiliation{%
  \institution{University of Southern California}
  \department{Thomas Lord Department of Computer Science}
  \streetaddress{4676 Admiralty Way (1001)}
  \city{Los Angeles}
  \state{CA}
  \postcode{90089}
  \country{USA}
}
\email{jinyiy@usc.edu}

\author{Luca Luceri}
\orcid{0000-0001-5267-7484}
\affiliation{%
  \institution{University of Southern California}
  \department{Information Sciences Institute}
  \streetaddress{4676 Admiralty Way (1001)}
  \city{Marina Del Rey}
  \state{CA}
  \postcode{90292}
  \country{USA}
}
\email{lluceri@isi.edu}

\author{Emilio Ferrara}
\orcid{0000-0002-1942-2831}
\affiliation{%
  \institution{University of Southern California}
  \department{Thomas Lord Department of Computer Science}
  \city{Los Angeles}
  \state{CA}
  \postcode{90089}
  \country{USA}
}
\email{emiliofe@usc.edu}

\renewcommand{\shortauthors}{Ye et al., 2025}

\begin{abstract}
Approximately 50\% of tweets in $\mathbb{X}$'s user timelines are personalized recommendations from accounts they do not follow. This raises a critical question: What political content are users exposed to beyond their established networks, and what implications does this have for democratic discourse online? 
In this paper, we present a six-week audit of $\mathbb{X}$'s algorithmic content recommendations during the 2024 U.S. Presidential Election by deploying 120 sock-puppet monitoring accounts to capture tweets from their personalized ``For You'' timelines. Our objective is to quantify \textit{out-of-network} content exposure for right- and left-leaning user profiles and assess any potential inequalities and biases in political exposure. Our findings indicate that $\mathbb{X}$'s algorithm skews exposure toward a few high-popularity accounts across all users, with right-leaning users experiencing the highest level of exposure inequality. Both left- and right-leaning users encounter amplified exposure to accounts aligned with their own political views and reduced exposure to opposing viewpoints. Additionally, we observe that new accounts experience a right-leaning bias in exposure within their default timelines. Our work contributes to understanding how content recommendation systems may induce and reinforce biases while exacerbating vulnerabilities among politically polarized user groups. We underscore the importance of transparency-aware algorithms in addressing critical issues such as safeguarding election integrity and fostering a more informed digital public sphere.
\end{abstract}


  

\maketitle

\sloppy
\section{Introduction}
During the 2024 U.S. Presidential Election, social media platforms like $\mathbb{X}$ (formerly Twitter) play a pivotal role as hubs for political information and public discourse. However, the information users encounter on $\mathbb{X}$ is increasingly curated by algorithmic recommendation systems that personalize content in their ``For You'' timelines. As of this writing, $\mathbb{X}$'s ``For You'' timeline typically consists of 50\% \textit{in-network} tweets (i.e., from accounts a given user follows) and 50\% \textit{out-of-network} tweets (i.e., from accounts that user does not directly follow)---an increase from the 37\%  \textit{out-of-network} tweets in 2023 \cite{bouchaud2023crowdsourced}. 

How does $\mathbb{X}$'s algorithm select relevant tweets from outside a user’s network? In 2023, Twitter partially open-sourced its recommendation algorithm, revealing that out-of-network recommendations are sourced through engagement and follow graphs, ranked by a neural network, and refined with heuristics and filters \cite{twitter_algorithm}. Despite the increasing prominence of out-of-network tweets in user timelines, much remains unknown about their composition and nature. While prior research has demonstrated amplification of certain political groups and media sources within users' \textit{in-network} tweets, the extent to which such biases extend to \textit{out-of-network} recommendations is unclear. In contexts like the 2024 U.S. Election, examining this issue is essential to understanding how algorithms shape the consumption of online political content and influence users' perspectives.

Research on $\mathbb{X}$'s algorithmic auditing faces a critical challenge in analyzing out-of-network content: While many studies assess amplification by comparing personalized timelines with reverse-chronological timelines as a baseline---where tweets appear in the order they were posted without algorithmic effects \cite{bartley2021auditing, bartley2023evaluating, huszar2022algorithmic, bouchaud2023crowdsourced, wang2024lower}, out-of-network tweets lack a reverse-chronological baseline as users do not follow the authors of those tweets, making it challenging to quantitatively measure exposure bias. To address this limitation, we utilize a ``sock-puppet audit,'' a study design that deploys artificial user accounts with controlled features to systematically capture and analyze platform recommendations \cite{bartley2021auditing, chen2020neutral, bandy2021more, hosseinmardi2024causally}. Specifically, we introduce a self-constructed baseline using accounts that follow a politically balanced social media diet, enabling direct comparisons with other manually-created partisan user accounts. This approach is particularly well-suited to studying out-of-network exposure patterns because it allows us to observe algorithmic behavior without the interference of the variations of user behaviors or connections. Furthermore, previous sock-puppet audits on $\mathbb{X}$ are often constrained by small sample sizes (fewer than 10 accounts) and restricted tweet collection in terms of both quantity and frequency \cite{bandy2021more, bartley2021auditing, chen2020neutral}, limiting the generalization and robustness of their findings. Our audit aim to implement a more comprehensive data collection strategy, enabling us to systematically observe and analyze algorithmic behaviors on a much larger scale.

\paragraph{Contribution of this work}
In this study, we deploy 120 sock-puppet accounts distributed into four groups across the political spectrum---left-leaning, right-leaning, balanced, and neutral---collecting a robust dataset of over 9 million tweets over six weeks from October to November 2024. Within this framework, we systematically evaluate potential exposure biases, such as popularity bias and algorithmic (de-)amplification across account groups.

The contributions of this work can be summarized as follows:
\begin{itemize}
    \item \textbf{We quantify algorithmic exposure to out-of-network content} for users with varying political alignments during the 2024 U.S. Election through a sock-puppet audit of $\mathbb{X}$'s personalized timelines.
    
    \item \textbf{We propose a methodology for evaluating out-of-network (political) exposure biases} by creating a baseline using politically balanced accounts.
    
\end{itemize}


We find that $\mathbb{X}$ skews exposure toward a few high-popularity accounts for all users, with right-leaning users experiencing the most inequality. Both left- and right-leaning users encounter amplified exposure to accounts aligned with their own political stance and reduced exposure to opposing viewpoints. Additionally, neutral accounts who do not follow anybody (akin to a newly-registered user account) show a default right-leaning bias in content exposure. Our findings reveal how content recommendation systems can influence and amplify biases, potentially increasing vulnerabilities within politically polarized user groups. This work underscores the urgent need for transparent algorithms to safeguard the integrity of online discourse and the sovereignty of elections.

\section{Background \& Research Questions}
\subsection{Related Work}
The impact of algorithmic content curation on political discourse in social media has been a major focus of research and public debate. Previous studies consistently show that $\mathbb{X}$'s algorithm amplifies political biases and prioritizes high-engagement content, including emotionally charged, toxic, and low-credibility information \cite{huszar2022algorithmic, bouchaud2023crowdsourced, bandy2021more, chen2020neutral, bartley2021auditing, corsi2024evaluating}. Researchers have used methods including randomized experiments, sock-puppet audits, crowdsourced audits, and observational data to study $\mathbb{X}$'s algorithmic effects. Some have found that Twitter's algorithms tend to amplify content from right-leaning media sources and politicians more than their left-leaning counterparts \cite{huszar2022algorithmic, graham2024algorithmic_bias}. Other studies report increased exposure to ideologically aligned friends \cite{bouchaud2023crowdsourced, bandy2021more}, but decreased exposure to external links \cite{wang2024lower, bandy2021more}. Studies also observe increased low-credibility content in algorithmic timelines \cite{corsi2024evaluating}, with right-leaning users experiencing higher exposure to such content \cite{chen2020neutral}. Although algorithms are often flagged for promoting ideological bias and political polarization \cite{barbera2020social}, as observed on platforms like YouTube \cite{haroon2023auditing}, other analyses of $\mathbb{X}$ and YouTube suggest that its algorithm tends to push centrist content to partisan users \cite{chen2020neutral, hosseinmardi2024causally} and displays a more diverse political mix overall \cite{bouchaud2023crowdsourced, wang2024lower}.

Despite these insights, the existing literature has a key limitation. Algorithmic timelines consist of two distinct components: the reordering and filtering of in-network tweets and the rendering of out-of-network recommendations. While most studies treat the timeline as a unified entity, making it difficult to disentangle biases between these components, our study focuses explicitly on the latter---out-of-network recommendations---which has received little attention in prior research. Our focus is particularly relevant after Elon Musk's takeover of the platform, as subsequent changes to content moderation and algorithmic priorities \cite{twitter_algorithm} may have heightened the impact of out-of-network recommendations on user experiences. In what follows, we outline the algorithmic biases under investigation and introduce our research questions (RQs).

\subsection{Exposure Inequality}
One significant aspect of algorithmic biases on social media is popularity bias \cite{nikolov2019quantifying}. Algorithms often tend to amplify content from certain users over others, creating inequalities in exposure \cite{bartley2023evaluating}. For instance, Twitter's ranking algorithm employs a \textasciitilde48M parameter neural network, which uses thousands of features to score each tweet based on engagement probabilities, prioritizing content with higher likelihoods of interaction in users' feeds \cite{twitter_algorithm}. Previous research has shown that popularity biases can lead to a skew in the visibility of tweets when comparing personalized feeds with reverse-chronological ones, and that users are disproportionately exposed to friends' tweets \cite{bartley2023evaluating, bartley2021auditing}. Yet, it remains unclear whether exposure inequalities extend beyond friends to include a broader set of recommended users. Specifically, we pose the following RQ: 
\begin{quote} 
\textit{\textbf{RQ1:} To what extent do personalized recommendations in $\mathbb{X}$ exhibit exposure inequality among users, and how do these inequalities differ based on political leanings?} 
\end{quote}

\subsection{Out-of-Network (De-)Amplification}
Another key dimension of bias is ideological bias, particularly its relationship with algorithmic (de-)amplification and selective exposure to political content.
Selective exposure is a psychological concept that refers to the tendency of individuals to prefer information that aligns with their pre-existing beliefs, attitudes, or preferences, while avoiding information that contradicts them \cite{guess2018selective}. 
Algorithms on social media platforms can amplify this effect by recommending content similar to what users already prefer or agree with, reinforcing selective exposure through personalization \cite{knudsen2023modeling}. 
Existing research has produced mixed findings on this issue. On one hand, \citet{bakshy2015exposure} report considerable cross-cutting exposure on Facebook, and \citet{wang2024lower} find that Twitter/$\mathbb{X}$ provides higher-quality and less ideologically congruent news curation. On the other hand, \citet{haroon2023auditing} trained sock puppets to represent five ideological positions ranging from left to right, and found that YouTube's algorithm consistently promotes ideologically aligned content to partisan users. Given the inconsistency in findings and our focus on algorithmic recommendations, we seek to address the following RQ:
\begin{quote}
\textit{\textbf{RQ2:} Which out-of-network users are (de-)amplified in the timelines of left- and right-leaning accounts compared to balanced accounts?}
\end{quote}

In the next sections, we outline our experimental setup, data collection process, and methodology designed to address our RQs.

\section{Methods}

\subsection{Experimental Setup}
We create 120 sock-puppet accounts on $\mathbb{X}$ divided into four groups based on their political leaning: 30 neutral accounts (default setting, following no one), 30 left-leaning accounts, 30 right-leaning accounts, and 30 balanced accounts. One general concern about sock-puppet audits is ecological validity---whether artificial accounts accurately represent real user behavior and interactions \cite{wang2024lower, bandy2021more, bandy2021problematic}. However, since we are auditing political biases in algorithmic recommendations, which can be influenced by user engagement and community affiliation, it is crucial to control user behavior as much as possible. Previous studies often deploy bots that mimic real-world content consumption by replicating real user follows as ``preset'' \cite{bandy2021more, hosseinmardi2024causally}. However, real users often follow diverse, non-political accounts, which could confound our focus on political contents. To address this, we limit our sock-puppet accounts to follow exclusively media, political figures, and entities.

We define the orientation of these sock-puppets based on the accounts they follow. To categorize the political alignment of accounts to follow, we use the AllSides Media Bias Chart,\footnote{\textit{AllSides Media Bias Chart} \url{https://www.allsides.com/media-bias/media-bias-chart}} which rates news sources on a spectrum from left to right based on their political bias. Each left-leaning and right-leaning account follows 10 media outlets, including seven outlets with a moderate (center-left or center-right) bias and three with a stronger (left or right) bias, as defined by the AllSides' chart. This selection ensures that these accounts represent a realistic mix of moderately and strongly aligned sources, enhancing the accuracy of our analysis of political exposure. Additionally, left-leaning accounts follow key Democratic figures and entities (Kamala Harris, Tim Walz, House Democrats, and Senate Democrats), while right-leaning accounts follow their Republican counterparts (Donald Trump, JD Vance, House Republicans, and Senate Republicans). Balanced accounts, designed to reflect a centrist perspective, follow five center-left and center-right media outlets and both presidential candidates from each major party. All media follows are randomly selected from the respective groups in the media bias chart, ensuring consistency with each group's intended alignment.

While some studies, particularly on YouTube, allow bots to interact with algorithms (e.g., following recommendations to study radicalization \cite{hosseinmardi2024causally, haroon2023auditing}), we refrain from inducing interactions in the current study for several reasons. First, interactions can create feedback loops that distort the algorithm's outputs, making it difficult to isolate baseline biases. Second, interaction-based designs complicate comparisons across accounts, as partisan accounts might engage differently with recommendations, introducing variability that is hard to standardize. Third, our focus is on measuring how algorithms recommend political content based on baseline configurations, such as predefined follows. Unlike radicalization studies, which examine user-algorithm feedback, our goal is to capture inherent biases in the recommendation system, best analyzed without user interactions.

We also take efforts to mitigate bias in the design of sock-puppet accounts. According to the $\mathbb{X}$ platform, each account was required to select at least three interests at the time of creation. We use a program to select these interest randomly, alongside random birthdates between 1990 and 1999. To further randomize account attributes and mitigate location-based biases in recommendations, a VPN was used during data collection. These steps ensured consistent and relatively unbiased data capture while adhering to platform constraints.

\subsection{Data Collection}
We develop a timeline crawler to systematically collect tweets recommended to different types of user profiles in $\mathbb{X}$'s ``For You'' timeline. 
The timelines for each account are collected four times daily, yielding approximately 500–700 tweets per session, or about 2,000–3,000 tweets per account per day, within the limits that $\mathbb{X}$'s terms of service impose on new, non-premium accounts. The choice of four daily scraping sessions was made to capture the variability in recommendations throughout the day, as the content recommended by $\mathbb{X}$’s algorithm can shift based on temporal factors like recent events or trending topics. It provides a more comprehensive picture of the algorithmic exposure that users might experience. Data collection spanned from October 2, 2024, one month before the election, to November 19, 2024, two weeks after the election, yielding a dataset of 9.79 million tweets. Figure \ref{fig:data_collection_overview} in the Appendix display the number of active accounts and the total tweets collected daily.


Table \ref{tab:dataset_stats} provides an overview of the statistics for the collected tweet dataset across different account types. It shows the average proportion of out-of-network tweets that each account type encounters, with neutral accounts seeing exclusively out-of-network content, while the other accounts have 55\%-63\% of their timelines composed of out-of-network tweets. Additionally, it details the average proportions of retweets, quoted tweets, and promoted tweets observed by each account type.

\begin{table*}[t]
\centering
\caption{Statistics of the collected dataset (mean values with standard deviations)}
\label{tab:dataset_stats}
\begin{tabular}{lcccc}
\toprule
\textbf{Statistic} & \textbf{Neutral} & \textbf{Left} & \textbf{Right} & \textbf{Balanced} \\
\midrule
Out-of-network tweet & 100\% & 59.23\% (7.45) & 55.88\% (6.66) & 62.27\% (5.70) \\
Retweet & 0.15\% (0.66) & 2.93\% (1.07) & 2.54\% (1.35) & 2.41\% (1.62) \\
Quoted tweet & 1.37\% (2.32) & 8.67\% (2.33) & 12.65\% (2.90) & 11.98\% (2.00) \\
Promoted tweet & 1.36\% (1.65) & 7.43\% (0.60) & 7.21\% (0.68) & 7.84\% (1.39) \\
\bottomrule
\end{tabular}
\end{table*}

\subsection{Exposure Evaluation Metric}
To measure a user's exposure within a timeline, we introduce a metric called ``weighted occurrence per 1,000 tweets,'' defined as the number of times a user's tweets appear per 1,000 tweets in the timeline, weighted by each tweet's visibility according to its rank. This adjustment gives more weight to tweets that appear earlier in one's timeline, as those tweets are also the more likely to be seen by a user and are known to generate more engagements \cite{kang2015vip}. For each $\mathbb{X}$ user whose tweet appears in the personalized timelines of our monitoring accounts, the ``weighted occurrence per 1,000 tweets'' metric is mathematically expressed as: 
$$
\text{Weighted Occurrence Per 1K Tweets} = \frac{1}{N} \sum_{i=1}^{n} p_{i} \cdot 1000,
$$

where $p_{i}$ is the probability of exposure related to a specific tweet, $n$ denotes the total number of times the user's tweets appear in the monitoring account's timeline, and $N$ is the aggregate count of tweets in all timelines collected for the monitoring account.

The probability of exposure, $p_{i}$, represents the estimated likelihood that a tweet is seen by a real user. Items near the top of a user's social media feed are more visible and thus more likely to be viewed. Following prior work on modeling collective attention on social media \cite{wu2007novelty, li2020dynamics}, we employ an exponential decay function, $p\left( r \right) = A \cdot e^{-\lambda r}$, to approximate the probability that a tweet at a given rank $r$ in a timeline will be seen. Each tweet in the sequence is assigned a weight that decreases gradually from 1 towards 0, representing the declining probability of user exposure as the tweet's position moves further down the timeline.

The parameters of the exponential decay function are informed by findings from studies on platforms like TikTok and YouTube \cite{guinaudeau2022fifteen}, which indicate that the top 20\% of an account's videos receive more than 70\% of the views. Using this as a reference, we assume that the top 20\% of tweets in a timeline similarly capture the majority (70\%) of user attention, and we calibrate our decay model accordingly. For instance, for a neutral account with an average timeline length of 500, the exponential decay function is defined as: 
\[
p_{\text{neutral}}\left( r \right) = 1.009 \cdot e^{-0.0120 \cdot r}.
\]

\subsection{Inequality Measure}
\paragraph{Gini Coefficient}
To measure whether exposure is evenly distributed among users or dominated by a few accounts, we employ the Gini coefficient, a widely used measure to quantify inequality \cite{bartley2023evaluating, farris2010gini}. The Gini coefficient ranges from 0 to 1, where 0 indicates perfect equality (all users have the same exposure) and 1 signifies maximum inequality (exposure is concentrated among a few accounts). In our specific case, the Gini coefficient $G$ is calculated as:
\[
G = \frac{\sum_{i=1}^{n} \sum_{j=1}^{n} |E_i - E_j|}{2n^2 \bar{E}},
\]
where \( E_i \) and \( E_j \) represent the exposure metrics---weighted occurrence per 1,000 tweets---of users \( i \) and \( j \) in a monitoring account's timeline, \( n \) is the total number of users, and \( \bar{E} \) is the mean exposure metric across all users. A higher Gini coefficient indicates greater inequality in exposure distribution, suggesting that a small number of users dominate exposure in the timeline, while a lower coefficient suggests a more even distribution among users. To complement this analysis, we use the Lorenz curve \cite{gastwirth1971general} as a visual representation of exposure inequality.

\subsection{Amplification Measure}
To assess the (de-)amplification of specific users in relation to left- and right-leaning monitoring accounts compared to a baseline constructed from balanced accounts, we introduce the ``mean amplification ratio,'' inspired by the work of \citet{huszar2022algorithmic} on algorithmic amplification.

The mean amplification ratio $a_{u}$ for a user 
$u$, take the example of left-leaning monitoring accounts, is defined by the formula:
\[
a_u = \left( \frac{\bar{E}_u^{\text{left}} + 1}{\bar{E}_u^{\text{balanced}} + 1} - 1 \right) \times 100\%,
\]

where:
\[
  \bar{E}_u^{\text{left}} = \frac{1}{|V_{\text{left}}|} \sum_{v \in V_{\text{left}}} E_{v,u},
\]

\[
  \bar{E}_u^{\text{balanced}} = \frac{1}{|V_{\text{balanced}}|} \sum_{v \in V_{\text{balanced}}} E_{v,u}.
\]

Here, \( V_{\text{left}} \) is the set of left-leaning accounts, and \( V_{\text{balanced}} \) is the set of balanced accounts. \( E_{v,u} \) denotes the weighted occurrence per 1,000 tweets of account \( u \) in the timelines of account \( v \). This amplification ratio quantifies the extent to which a user's exposure is increased or decreased when viewed by left-leaning monitoring accounts compared to the balanced baseline. A positive \textit{mean amplification ratio} indicates amplification, while a negative ratio indicates de-amplification. The calculation for right-leaning monitoring accounts follows a similar approach.

\section{Results}
\subsection{Out-of-Network Exposure Inequality Among Different Political Profiles (RQ1)}

RQ1 explores the extent to which personalized recommendations in $\mathbb{X}$ exhibit exposure inequality among users and how these inequalities vary between partisan accounts. To address this question, we use the Gini coefficient, a standard measure of inequality that quantifies disparities in exposure by calculating how concentrated exposure is across a set of users. Detailed descriptions of the Gini coefficient calculation and the exposure metric are provided in the Methods section. For each sock-puppet monitoring account, we compute its Gini coefficient with respect to all recommended users in that account's timelines. 

\begin{figure}[t]
    \centering
    \includegraphics[width=\linewidth]{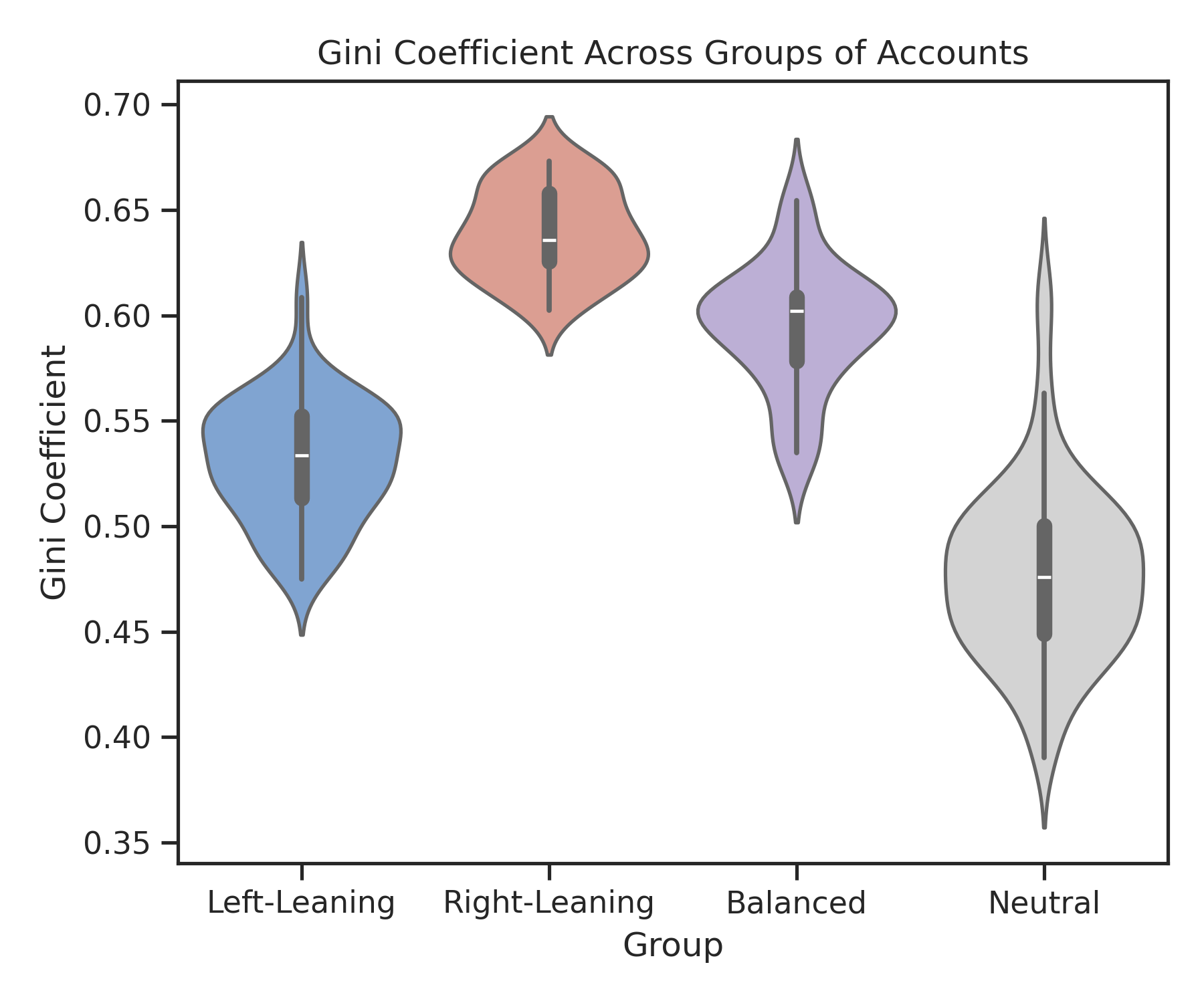}
    \caption{Distribution of Gini coefficient across different groups of accounts. Significant disparities are found in all pairwise comparisons (Mann-Whitney U test: $p$ < 0.001), with right-leaning users experiencing the highest out-of-network exposure inequality.}
    \Description{Gini coefficient violin plot}
    \label{fig:gini_coefficient}
\end{figure}

\begin{figure}[t]
    \centering
    \includegraphics[width=\linewidth]{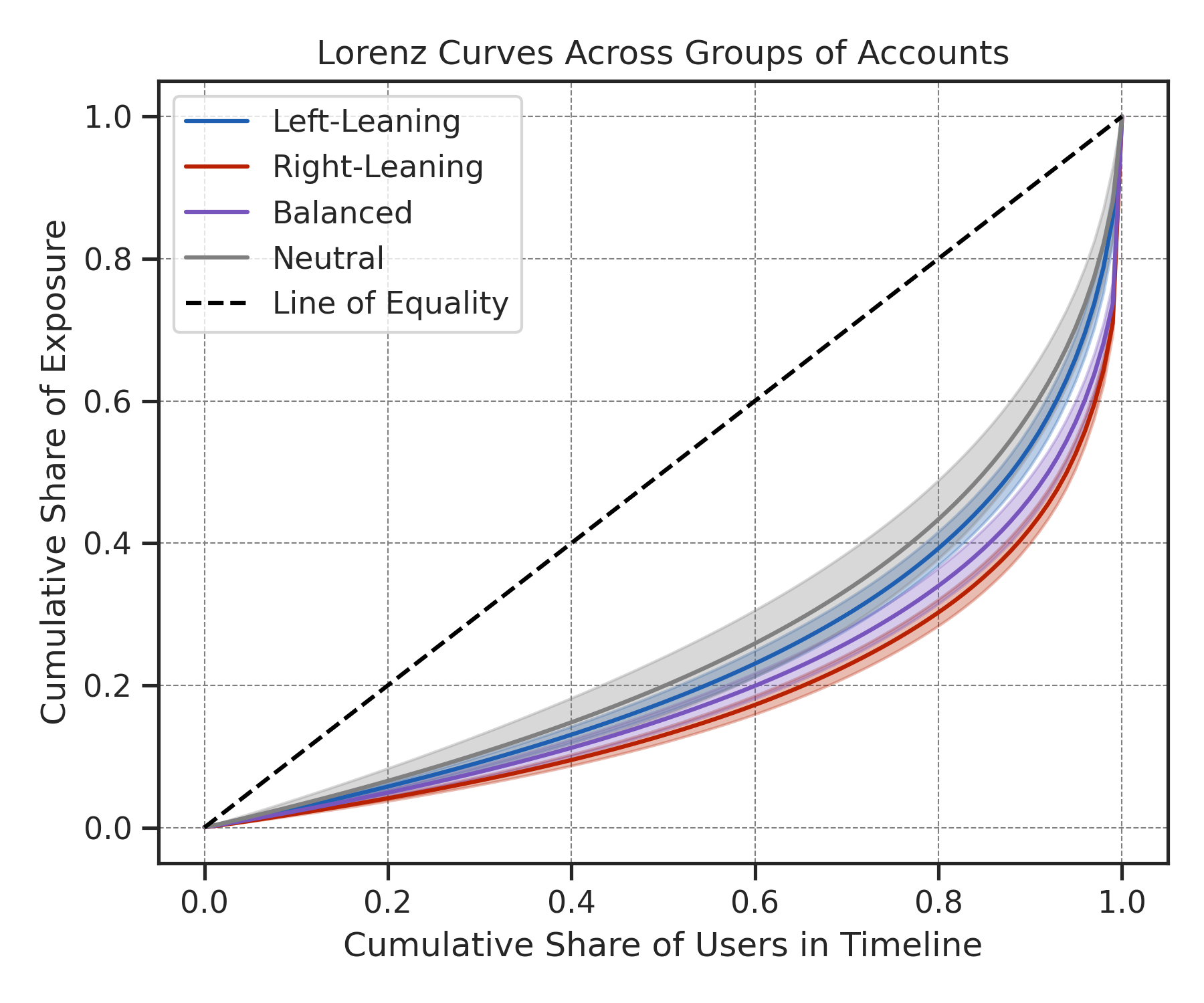}
    \caption{Lorenz curves for different groups of accounts. Each curve represents the average Lorenz curve for all accounts in the group, with error bars indicating the standard deviations at each cumulative point.}
    \Description{Lorenz curve cumulative plot}
    \label{fig:lorenz_curve}
\end{figure}


Figure \ref{fig:gini_coefficient} presents the distribution of Gini coefficients across different account groups: Left-Leaning, Right-Leaning, Balanced, and Neutral. The average Gini coefficient across all groups exceeds 0.45, which suggests a moderate to high level of inequality in exposure on the $\mathbb{X}$ platform. Similarly, as shown in Figure \ref{fig:lorenz_curve}, the Lorenz curves for all account groups deviate substantially from the line of equality (dashed black line). The greater the curvature of the Lorenz curve, the higher the inequality in exposure. This indicates that algorithmic exposure is concentrated among certain users rather than evenly distributed.

\begin{figure*}[t]
    \centering
    \includegraphics[width=\textwidth]{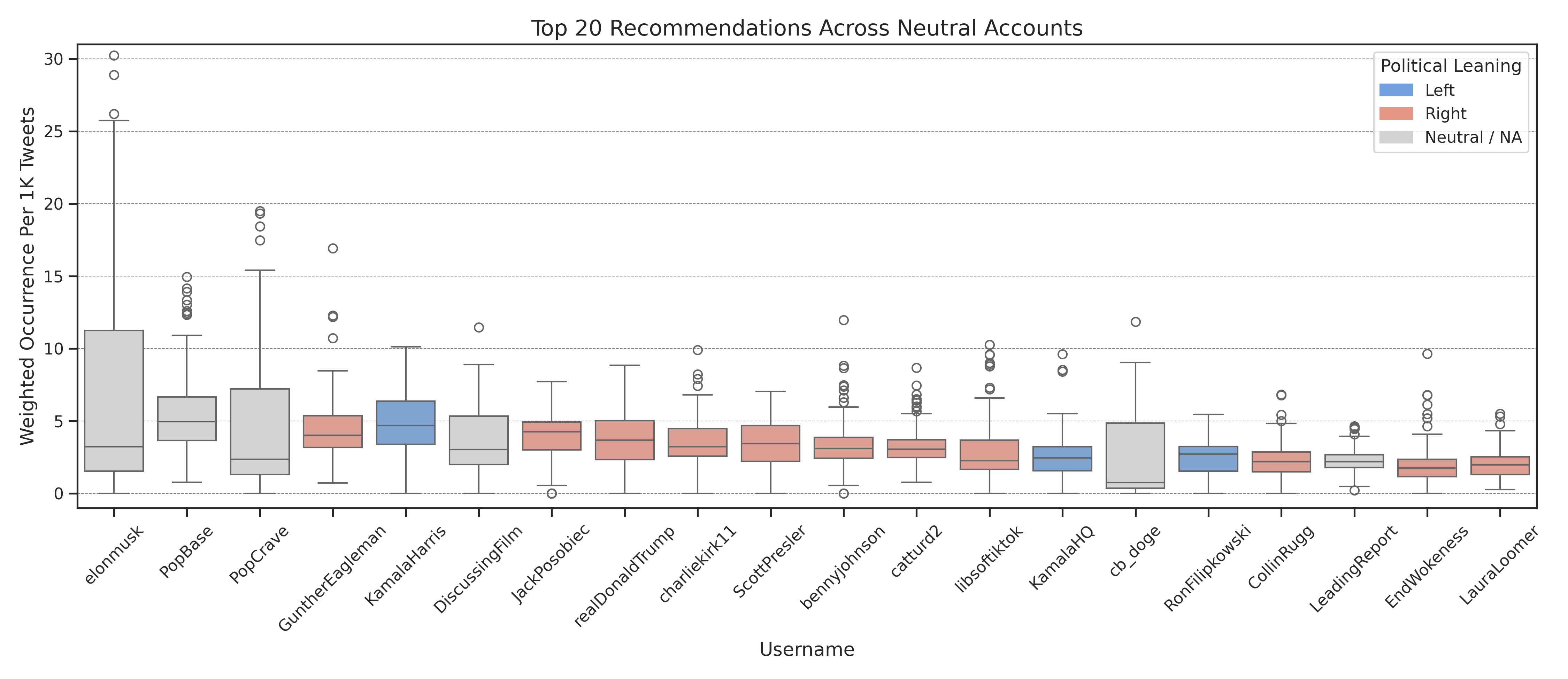}
    \caption{Top 20 recommended users for neutral accounts, ranked by their average weighted occurrence per 1,000 tweets. Each box in the boxplot shows the distribution of exposure across all neutral accounts, with red and blue colors indicating right- and left-leaning users, respectively. The figure suggests that right-leaning users are more frequently recommended than left-leaning users in the algorithm's out-of-network recommendations for neutral accounts.}
    \Description{Top 20 recommendations neutral}
    \label{fig:top20_neutral}
\end{figure*}

Notably, right-leaning users experience the highest exposure inequality, followed by balanced and left-leaning users. The Mann-Whitney U test reveals that the differences in Gini coefficients between all pairs of groups are significant at the 0.001 level, underscoring meaningful disparities in exposure inequality across these groups. This suggests that the algorithm's out-of-network tweet recommendations for right-leaning users are more centralized, reflecting a stronger popularity bias, where a few users dominate exposure. In contrast, neutral users---who do not follow anyone---receive the most diverse recommendations, potentially due to \textit{algorithmic cold start}, i.e., the absence of information about user preferences that typically informs recommendations \cite{yuan2023user}.

Our findings are significant when compared to previous studies that report Gini coefficients of approximately 0.6–0.7 for inequality in exposure to friends' tweets \cite{bartley2021auditing}. This suggests that even beyond the friend network, exposure inequality remains at a similar level, indicating that the platform's algorithm amplifies certain accounts both within and outside of users' direct networks. 

Now that we understand that out-of-network exposures are skewed toward certain users, an important question arises: \textit{Who are these users?} Here, we are particularly interested in neutral accounts, which provide an unbiased look at the algorithm's default behavior. Since neutral accounts are critical for detecting bias, we took particular care in their setup to ensure neutrality. Neutral accounts follow no other accounts and, therefore, receive exclusively out-of-network recommendations. This configuration limits any bias that could arise from following choices, aiming to capture a baseline view of how the algorithm behaves when no user preferences are specified. However, it is worth noting that certain factors, such as $\mathbb{X}$'s default settings or trending topics, could still introduce slight biases into these recommendations.

Figure \ref{fig:top20_neutral} displays the top 20 recommended users for \textit{neutral} accounts, ranked by their weighted occurrence per 1,000 tweets. Each box in the boxplot represents the distribution of this exposure metric across all neutral accounts. Boxes are colored red or blue to indicate whether the user is right- or left-leaning, based on publicly available data, including $\mathbb{X}$ user profile descriptions and external sources such as Wikipedia. A user's political stance is classified as left- or right-leaning if they are affiliated with a political party or a media outlet with a recognized ideological alignment. A qualitative inspection reveals that right-leaning users appear more frequently among the top recommendations than left-leaning users. To quantify this difference, we use the ``weighted occurrences per 1,000 tweets'' metric: among the top 20 recommended users, right-leaning users make up 30.16\% of exposure, compared to 12.92\% for left-leaning users. This disparity persists as we expand the pool, with right-leaning users making up 35.26\% of exposure in the top 50 (versus 22.34\% for left-leaning users) and 31.39\% in the top 100 (versus 20.83\% for left-leaning users). 

Notably, \textit{balanced} accounts receive a roughly even mix of left- and right-leaning recommendations, whereas left- and right-leaning accounts predominantly receive recommendations from ideologically aligned users. In the Appendix, interested readers can find the top 20 recommendations for left-leaning, right-leaning, and balanced account groups, highlighting the most amplified users within each account category. A detailed table describing these users' public information is also provided in the Appendix.

\subsection{Differential (De-)Amplification of Political Content Among Partisan Accounts (RQ2)}

\begin{figure*}
    \centering
    \begin{subfigure}{\textwidth}
        \centering
        \includegraphics[width=\textwidth]{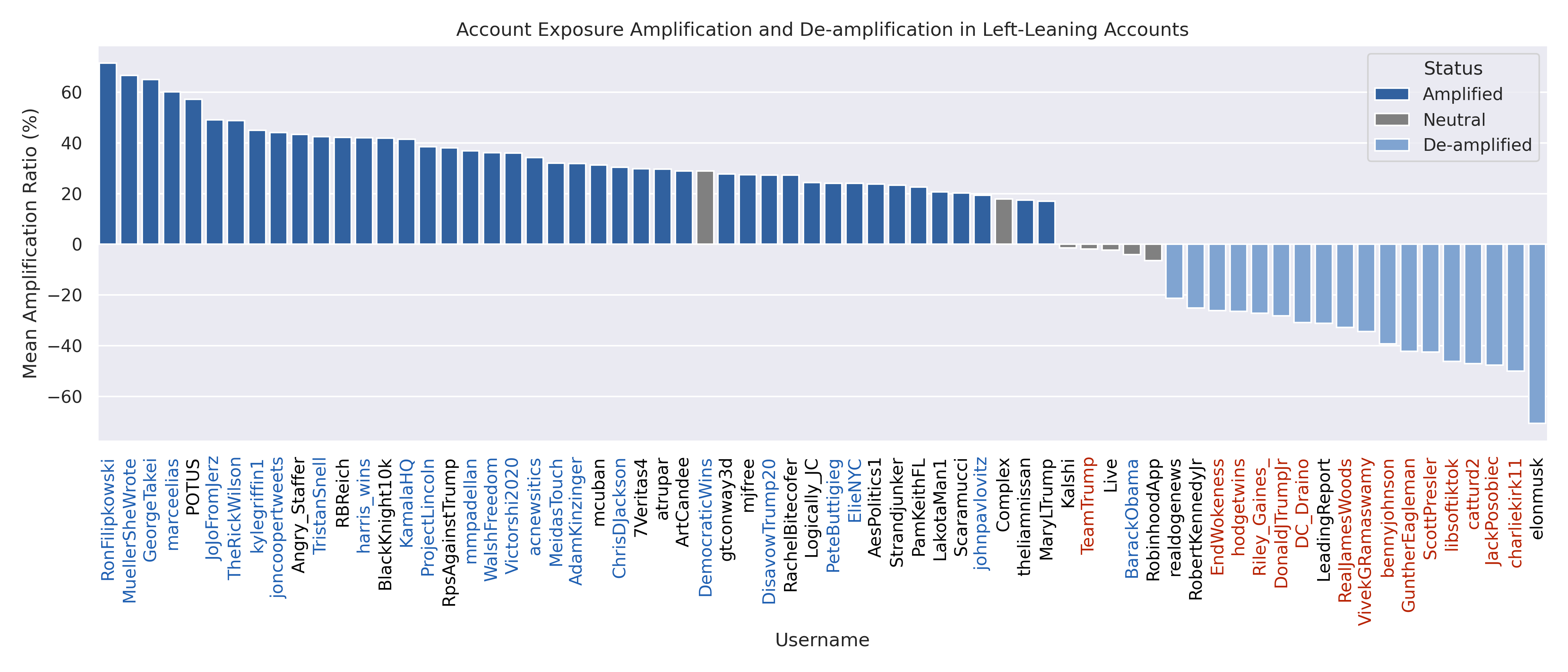}
        \label{fig:top50amplify_left}
        \Description{amplified_accounts_left_50}
    \end{subfigure}
    

    \begin{subfigure}{\textwidth}
        \centering
        \includegraphics[width=\textwidth]{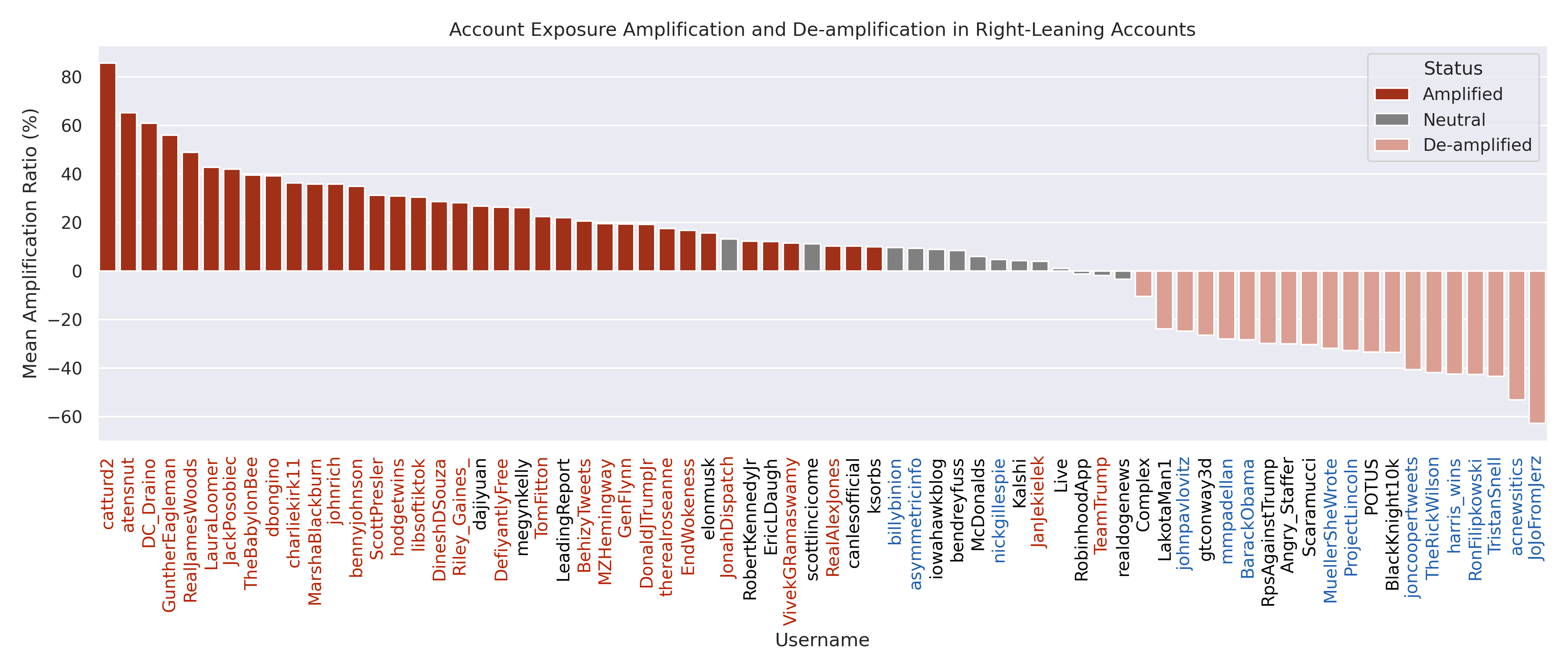}
        \label{fig:top50amplify_right}
        \Description{amplified_accounts_right_50}
    \end{subfigure}
    
    \caption{Amplification ratio of the top 50 recommended users in left-leaning (top) and right-leaning (bottom) accounts, compared to the baseline of balanced accounts. Colored bars indicate a significant difference in exposure metrics (weighted occurrence per 1,000 tweets) between the groups at the 0.05 significance level (using the Mann-Whitney U test), while gray bars indicate no significant difference. Usernames are displayed in blue (left-leaning) or red (right-leaning) based on their political stance, according to publicly available data.}
    \label{fig:top50amplify}
\end{figure*}

To address RQ2 and evaluate the amplification of certain users in partisan accounts' timelines, we introduce the ``mean amplification ratio'' metric inspired by \citet{huszar2022algorithmic}, as detailed in the Methods section. Figure \ref{fig:top50amplify} shows the amplification ratio of the top 50 recommended users in left-leaning and right-leaning accounts, compared to a baseline observed in politically balanced accounts' timelines. Colored bars indicate a significant difference in exposure metrics (weighted occurrence per 1,000 tweets) between groups at the 0.05 significance level (using the Mann-Whitney U test), while gray bars indicate no significant difference. 

A qualitative inspection reveals that left-leaning sock-puppet accounts tend to see left-leaning users amplified, and right-leaning users de-amplified, with the opposite pattern observed for right-leaning accounts. For instance, in left-leaning accounts, the top three amplified users are \emph{Ron Filipkowski} (a former federal prosecutor known for his criticisms of conservative figures), \emph{Mueller, She Wrote} (a political commentary and investigative journalism account with a liberal stance), and \emph{George Takei} (an American actor, author and Democrat activist). In contrast, the most de-amplified accounts are \emph{Elon Musk} (CEO of Twitter/$\mathbb{X}$, who has recently shared conservative viewpoints), \emph{Charlie Kirk} (a conservative political activist), and \emph{Jack Posobiec} (a right-wing media personality and political activist). This suggests that left-leaning timelines prioritize left-aligned figures while downplaying right-leaning accounts.

On the contrary, for right-leaning accounts, the top three accounts with the highest amplification in right-leaning timelines are \emph{catturd2} (a right-wing influencer known for political satire), \emph{atensnut} (a conservative commentator), and \emph{DC\_Draino} (Rogan O'Handley, a right-wing political commentator). Conversely, the most de-amplified accounts are \emph{JoJoFromJerz} (a left-leaning political influencer), \emph{acnewsitics} (a liberal-leaning news commentator), and \emph{Tristan Snell} (a pro-democrat lawyer and legal commentator). This pattern highlights the algorithm's tendency to amplify conservative figures more heavily in right-leaning timelines while reducing exposure to left-leaning accounts.

To further illustrate this trend, usernames are displayed in blue (left-leaning) or red (right-leaning) based on their political stance, which is inferred from publicly available data (may be subject to inaccuracies or changes over time). As shown in Figure \ref{fig:top50amplify}, top liberal and conservative voices are amplified more than 50\% above baseline for left- and right-leaning users, respectively. Given that our sock-puppet accounts only follow a few moderately partisan media and politicians, it suggests that once a new user begins following a few partisan accounts, their algorithmic recommendations quickly become filled with like-minded voices.

Interestingly, we observe that amplified users in left-leaning group experience a slightly higher magnitude of amplification compared to those in right-leaning group ($M_{left}=36.76\%$, $M_{right}=30.29\%$, Mann-Whitney U $p < 0.05$). However, there are no significant differences in the extent of de-amplification between the two groups.

\section{Discussion \& Conclusions}
In this study, we present a six-week audit of algorithmic recommendations on $\mathbb{X}$'s ``For You'' timelines during the course of the 2024 U.S. Election. Using 120 sock-puppet accounts with left-leaning, right-leaning, balanced, and neutral political orientations, we observe that $\mathbb{X}$ skews exposure toward a select few high-popularity accounts for all users, with right-leaning users experiencing the highest level of inequality. Both left- and right-leaning users see amplified exposure to accounts aligned with their political stance, while exposure to opposing viewpoints is reduced. Additionally, analysis of neutral accounts with no follow activity reveals a default right-leaning bias in the platform's recommendations.

Our analysis of exposure inequality aligns with previous studies on algorithmic bias, which have reported similar amplification patterns within users' in-network content \cite{bartley2023evaluating}. However, our findings diverge from earlier research suggesting that personalized recommendations tend to be more centrist in political stance \cite{chen2020neutral, bouchaud2023crowdsourced, wang2024lower}. This discrepancy perhaps highlights a shift in $\mathbb{X}$'s algorithmic behavior, which might have moved away from promoting moderate content to reinforcing users' existing preferences more explicitly, especially in out-of-network recommendations. The results also add to the growing body of literature indicating that right-leaning accounts are often more prominently featured in algorithmic curation \cite{graham2024algorithmic_bias}, a trend seen here in the default bias toward right-leaning content for new or neutral accounts.

Another noteworthy observation is that, unlike prior research, which has primarily examined the amplification of tweets from media outlets \cite{wang2024lower} and political figures---especially elected legislators from major political parties \cite{huszar2022algorithmic}---our findings reveal that $\mathbb{X}$'s algorithm now also amplifies political commentators and influencers. This trend is most pronounced in the recommendations for neutral accounts, suggesting a shift in the algorithm's prioritization toward these types of voices. This shift could be influenced by recent claims that $\mathbb{X}$ prioritizes verified and paid subscription accounts\footnote{\textit{Tweet from Twitter/X CEO Elon Musk} 
\newline
\url{https://x.com/elonmusk/status/1650731557164818437?lang=en}}, potentially amplifying influencers who invest in these platform features. The prominence of these non-institutional voices in political content raises questions about the influence of individual commentators on public opinion, as their perspectives may carry a more personal or sensational tone compared to traditional media sources. Adding to the concerns, recent investigations uncovered state-sponsored foreign interference operations with financial backing of prominent political influencers.\footnote{\textit{Justice Department Disrupts Covert Russian Government-Sponsored Foreign Malign Influence Operation Targeting Audiences in the United States and Elsewhere}
\url{https://www.justice.gov/opa/pr/justice-department-disrupts-covert-russian-government-sponsored-foreign-malign-influence}} This underscores the need for further examination into how the recommendation algorithm’s priorities may shape political engagement and public discourse, especially during critical periods like an election year.

\paragraph{Implications and Future Research.}
Our research findings offer both theoretical and practical implications regarding the algorithm's influence on echo chambers and the design of transparency-aware content recommendation algorithms. The $\mathbb{X}$ algorithm's amplification of ideologically aligned out-of-network accounts, along with the reduced exposure to opposing viewpoints, suggests that algorithmic recommendations can reinforce echo chambers not just in the composition of social networks \cite{duskin2024echo} but also in the ideological framing of content circulating in the network. The increased prominence of non-institutional voices, such as verified political commentators and influencers, further exacerbates this issue by potentially introducing sensationalism and misinformation into these echo chambers \cite{de2024echo}. Additionally, the default right-leaning bias observed for neutral accounts suggests that new users are likely to encounter partisan content early in their engagement with the platform. This raises concerns about how early algorithmic shaping of timelines might influence political perspectives and preferences. Future research could address these concerns by 1) systematically comparing in-network and out-of-network exposure biases and 2) conducting user studies to investigate how algorithmically curated timelines influence political attitudes over time (see ``sociotechnical audit'' \cite{wang2024lower}).

The study also provides practical considerations for designing fair and transparent algorithms. Current recommendation systems appear to disproportionately amplify high-popularity accounts, creating inequality in exposure that may result in less personalized and miscalibrated recommendations for certain user groups \cite{abdollahpouri2020connection}. Fairness algorithms could address this by factoring in diversity constraints that balance the exposure of popular and less popular accounts. Platforms should enhance transparency around how algorithms prioritize specific users, particularly verified and paid subscription accounts for the Twitter/$\mathbb{X}$ scenario. Future research should focus on monitoring algorithmic shifts and developing transparency standards during high-stakes periods such as elections, public health crises, and social unrest, where equitable and informed public discourse is critical.

\paragraph{Limitations.}
We acknowledge several limitations in our research. 
First, the study is conducted during a six-week period leading up to the 2024 U.S. elections, a politically charged time that may differ from other contexts. This temporal limitation could affect the reproducibility of our results in less politically sensitive periods.
Second, we deliberately avoid inducing interactions between sock-puppet accounts and algorithmic recommendations to isolate baseline biases in the recommendation system. Although our sock-puppet auditing method ensures precise control over account behaviors, it does not account for personalization or the dynamics of user activity. Since the sock-puppet accounts do not engage with tweets (e.g., clicking, responding, or retweeting), our study does not capture the effects of user-algorithm interactions on political exposure bias.
Third, potential confounding factors, such as pre-selected interests, age, and location, may influence algorithmic recommendations for neutral accounts. While we took care to randomize these settings, their residual effects cannot be entirely ruled out. 
Forth, the use of balanced accounts as a baseline for measuring exposure biases may not fully capture the platform's broader algorithmic behavior across diverse user demographics or global political contexts, potentially limiting the generalizability of our findings. 


\paragraph{Ethical Statement.} Throughout our research process, we have adhered to stringent ethical standards to ensure the integrity and societal responsibility of our work. Our sock-puppet accounts were designed solely to follow media and public figures, observe, and collect data, without engaging in any interactions with real users on the $\mathbb{X}$ platform, thereby avoiding disruptions to other users' experiences. All personal-identifiable information utilized in this study pertains exclusively to public figures and is derived from publicly available data. Additionally, we have carefully considered the societal impacts of our research. To mitigate risks of overgeneralization or misinterpretation, we provide thorough contextual information and openly address the limitations of our findings. While acknowledging these potential risks, we posit that our work could contribute to the development of algorithmic transparency standards and inform platform responsibilities during politically sensitive periods in the long term.

\bibliographystyle{ACM-Reference-Format}
\bibliography{references}

\clearpage
\newpage
\appendix
\onecolumn
\section{Data Collection Details}
\begin{figure}[h]
    \centering
    \begin{subfigure}{0.95\linewidth}
        \centering
        \includegraphics[width=\linewidth]{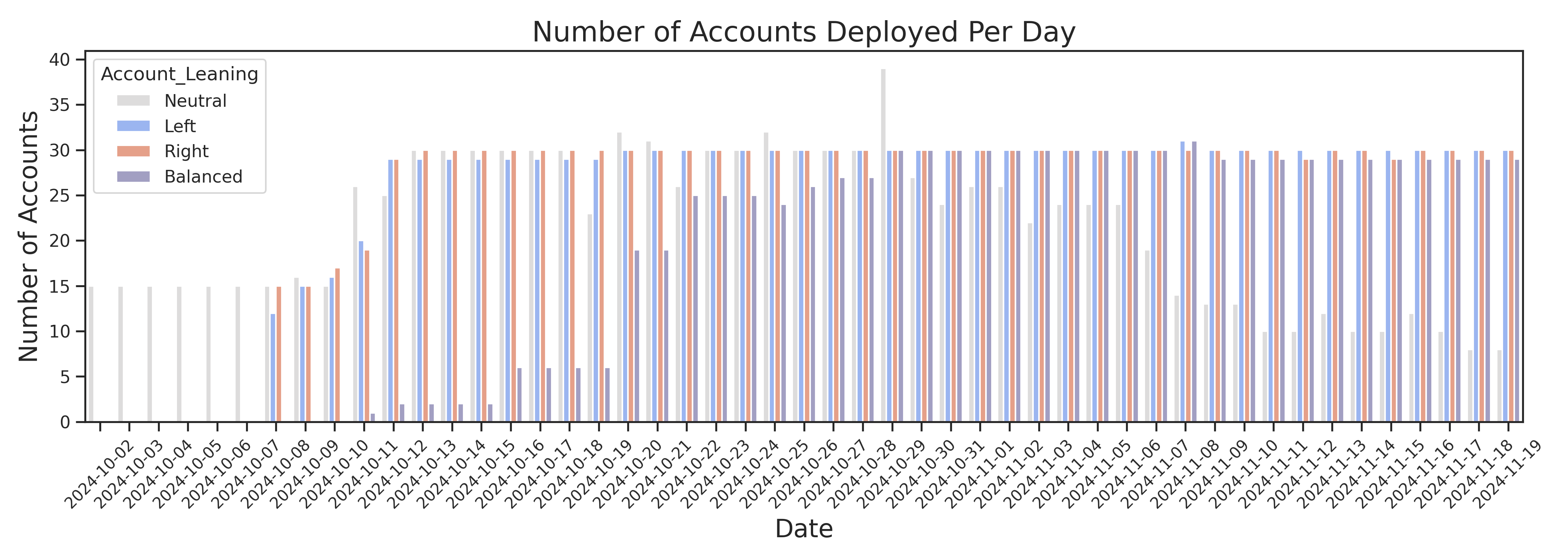}
        \label{fig:active_accounts}
        \Description{Number of bots}
    \end{subfigure}
    

    \begin{subfigure}{0.95\linewidth}
        \centering
        \includegraphics[width=\linewidth]{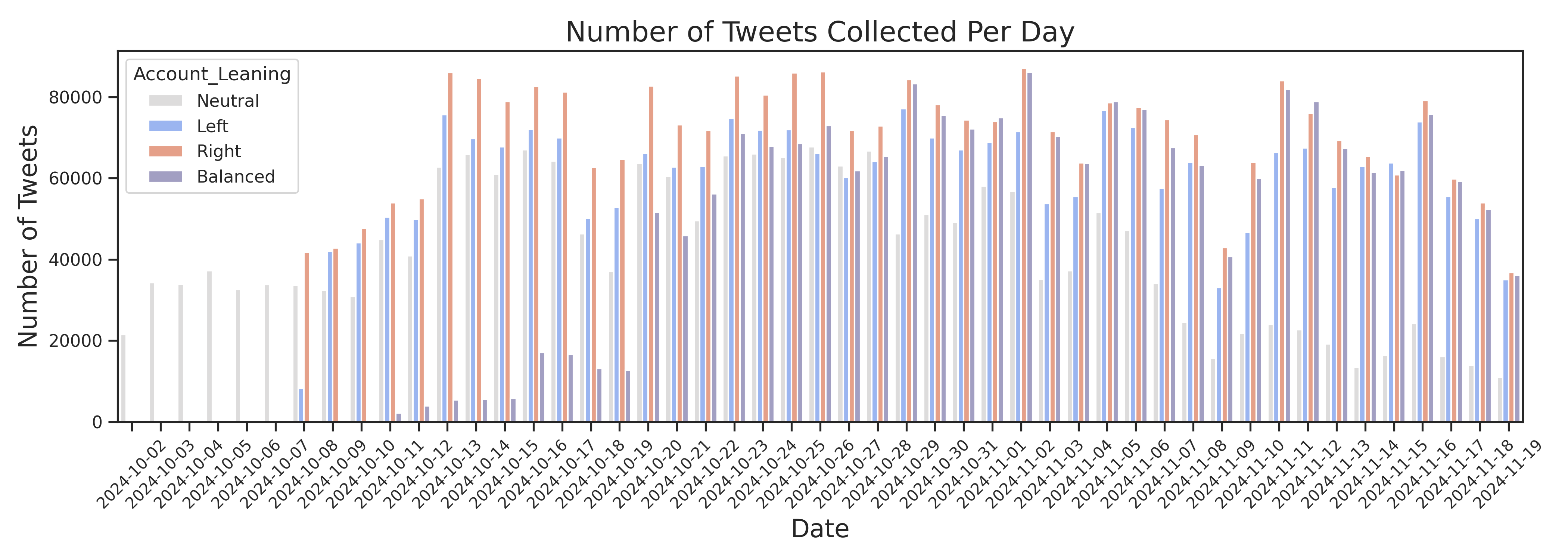}
        \label{fig:tweets_collected}
        \Description{Number of tweets}
    \end{subfigure}
    
    \caption{Overview of data collection: (a) Number of active accounts per day, and (b) Number of tweets collected per day.}
    \label{fig:data_collection_overview}
\end{figure}

Figure \ref{fig:data_collection_overview} display the number of active accounts and the total tweets collected daily. Data collection for neutral monitoring accounts began around October 2, 2024, and reached a stable deployment of approximately 30 active neutral accounts per day on October 11. Left-leaning, right-leaning, and balanced accounts began appearing consistently in the dataset around October 7, with each group reaching a stable count of about 30 active accounts per day shortly thereafter. Each neutral account receives approximately 500 tweets per session, while each left-leaning, right-leaning, and balanced account receives around 700 tweets per session. 

\paragraph{Technical Considerations for Neutral Accounts.}
Managing neutral accounts presented several challenges during data collection. For accounts that followed no users, $\mathbb{X}$ disabled the timeline after 7 days, requiring us to create additional neutral bots to maintain at least 30 active accounts daily. However, during the election period, the platform temporarily modified this restriction, disabling the timeline immediately after account creation for accounts that followed no users. Consequently, data collection was limited to approximately 10 older neutral accounts after November 5. 

\section{Top Recommended Users in Left-Leaning, Right-Leaning, and Balanced Accounts}
\begin{figure}[h]
    \centering
    \includegraphics[width=0.8\linewidth]{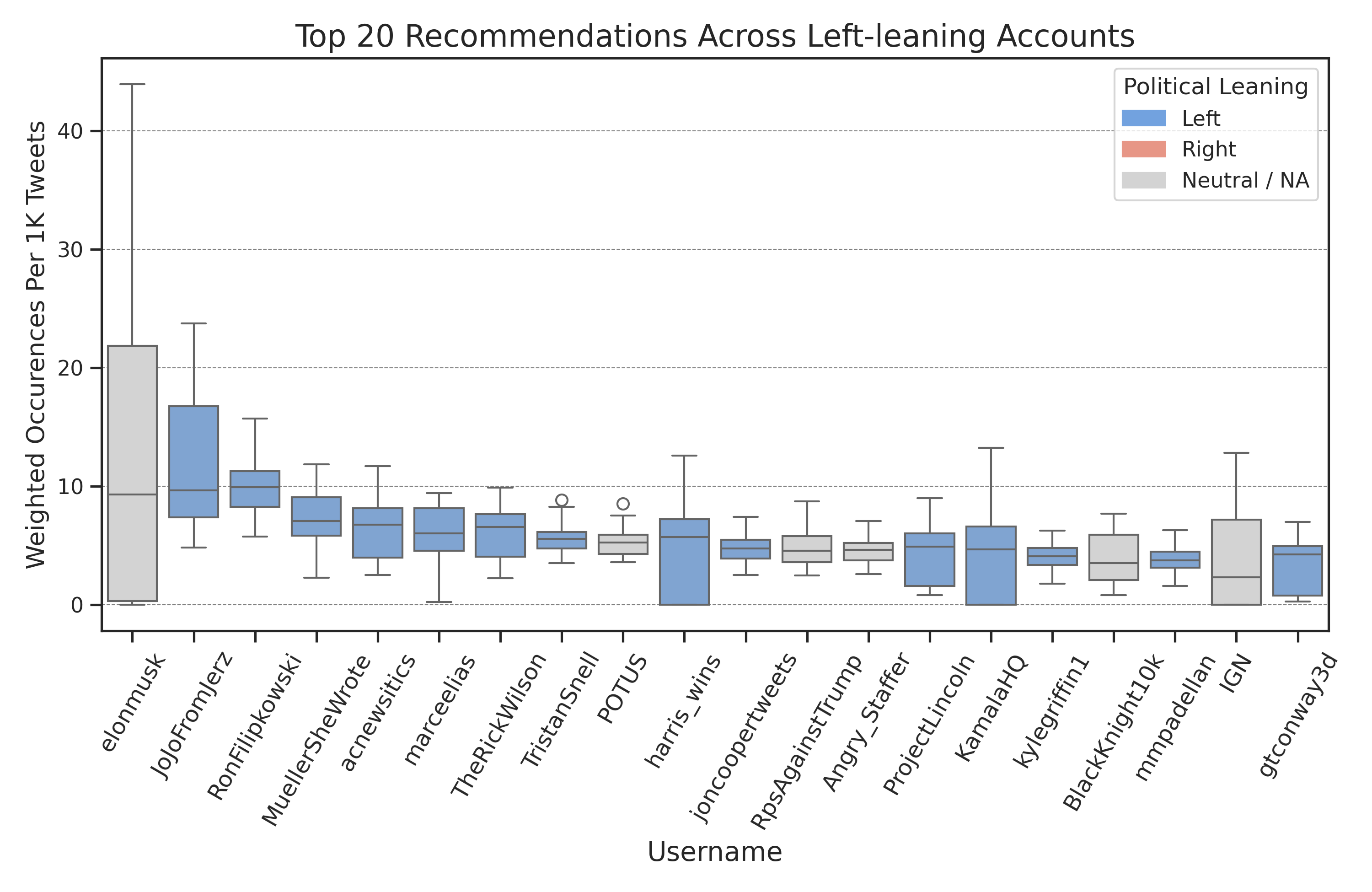}
    \caption{Top 20 recommended users in left-leaning accounts.}
    \label{fig:top20_left}
\end{figure}

\begin{figure}[t]
    \centering
    \includegraphics[width=0.8\linewidth]{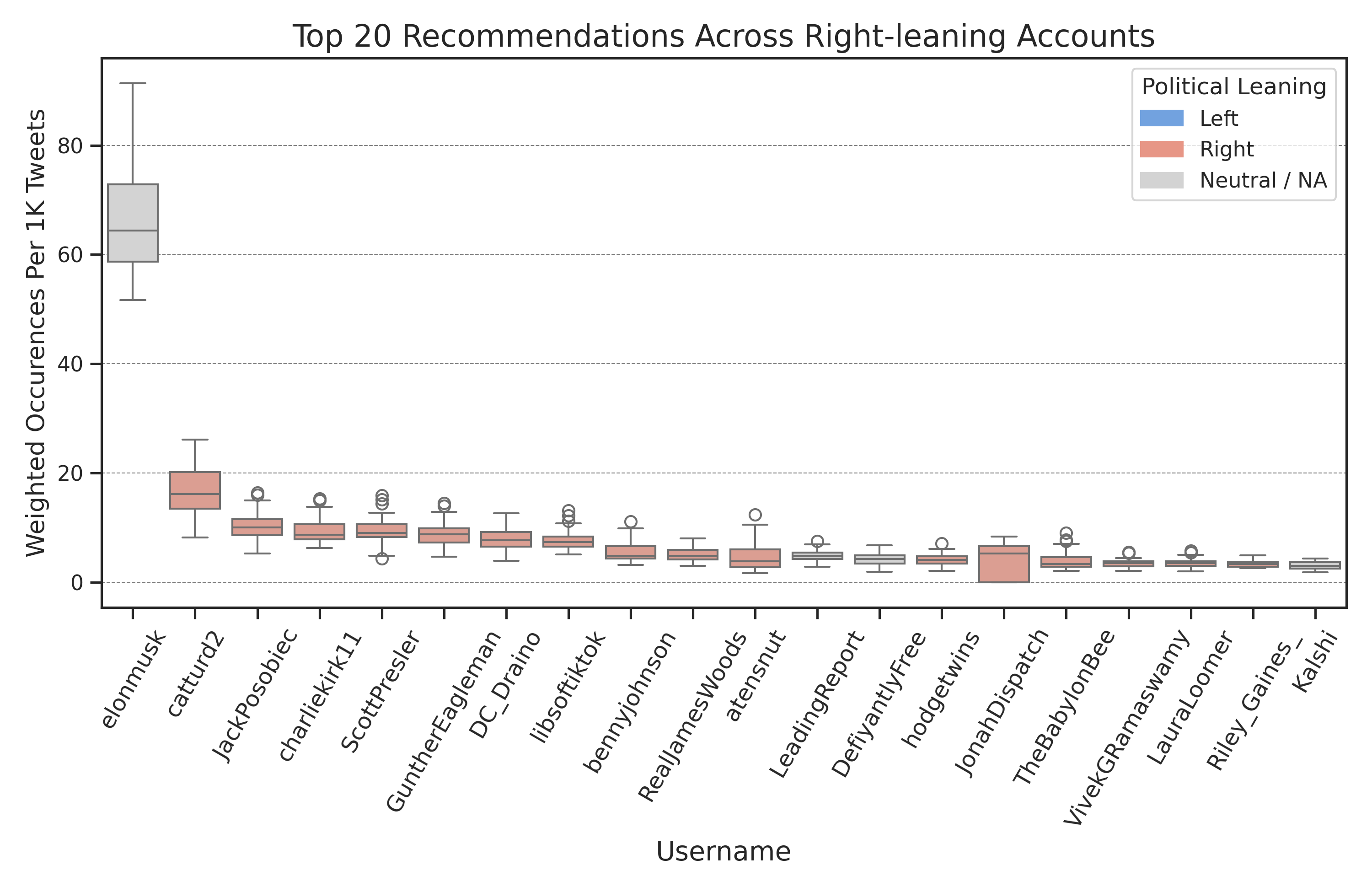}
    \caption{Top 20 recommended users in right-leaning accounts.}
    \label{fig:top20_right}
\end{figure}

\begin{figure}[t]
    \centering
    \includegraphics[width=0.8\linewidth]{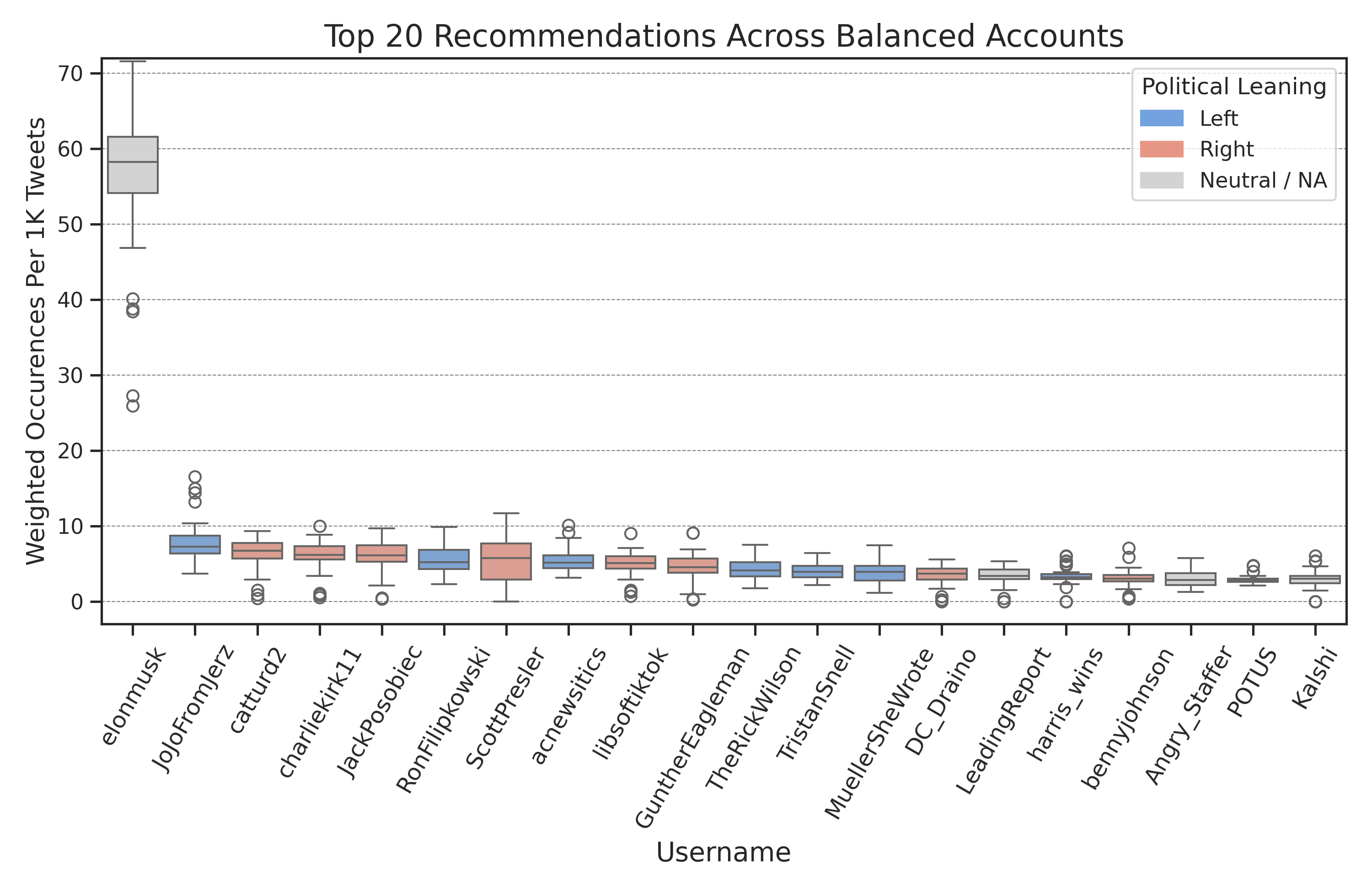}
    \caption{Top 20 recommended users in balanced accounts.}
    \label{fig:top20_balanced}
\end{figure}

Figure \ref{fig:top20_left}, figure \ref{fig:top20_right} and figure \ref{fig:top20_balanced} display the top 20 recommended users in left-leaning, right-leaning, and balanced accounts, ranked by their average weighted occurrence per 1,000 tweets. Each box in the boxplot represents the distribution of exposure across all accounts in each group, with red indicating right-leaning users and blue indicating left-leaning users. Political leanings of users are inferred based on publicly available data, which may be subject to inaccuracies or changes over time. Notably, balanced accounts receive a roughly even mix of left- and right-leaning recommendations, whereas left- and right-leaning accounts predominantly receive recommendations from ideologically aligned users.

\onecolumn
\section{Profile Information of the Top Recommended Users}
Table \ref{tab:top_profile} presents the profile information of the top 30 out-of-network recommendations across all account groups combined, sorted in descending order by the number of followers for each user. Note that, for left-leaning, right-leaning, and balanced accounts, top recommendations exclude media and politician accounts they already follow.

\small
\centering
\begin{longtable}{|c|l|p{2cm}|p{6cm}|c|}
\caption{Profile information of top out-of-network recommendations across all groups of accounts} \label{tab:top_profile}\\

\hline
\textbf{Index} & \textbf{Username} & \textbf{Screenname} & \textbf{Profile Description on $\mathbb{X}$} & \textbf{\# of Followers} \\
\hline
\endfirsthead

\hline
\textbf{Index} & \textbf{Username} & \textbf{Screenname} & \textbf{Profile Description on $\mathbb{X}$} & \textbf{\# of Followers} \\
\hline
\endhead

\hline
\endfoot
1 & elonmusk & Elon Musk & Read @America to understand why I'm supporting Trump for President & 202742780 \\
\hline
2 & BarackObama & Barack Obama & Dad, husband, President, citizen. & 132026578 \\
\hline
3 & realDonaldTrump & Donald J. Trump & 45th President of the United States of America & 92023938 \\
\hline
4 & POTUS & President Biden & 46th President of the United States, husband to @FLOTUS, proud dad \& pop. & 36825725 \\
\hline
5 & KamalaHarris & Kamala Harris & Fighting for the people. Wife, Momala, Auntie. She/her. Official account is @VP. & 21259465 \\
\hline
6 & AdamSchefter & Adam Schefter & \begin{tabular}[c]{@{}l@{}}ESPN Senior NFL Insider. \\ Interview \& Podcast Requests: ESPNPR@espn.com\\ Host of the Adam Schefter Podcast\\ https://t.co/oz43ix5jZU\end{tabular} & 11319193 \\
\hline
7 & Live & Live &  & 9024660 \\
\hline
8 & mcuban & Mark Cuban & Dunking on the pharma industry with @costplusdrugs.com, the lowest prices on meds anywhere. check it out ! & 8959820 \\
\hline
9 & dbongino & Dan Bongino & Public Enemy \#1 & 5881895 \\
\hline
10 & historyinmemes & Historic Vids & Daily history lessons. Education through memes! & 5451424 \\
\hline
11 & AMAZlNGNATURE & Nature is Amazing & Animals  Nature  Discovery  & 4496570 \\
\hline
12 & RealJamesWoods & James Woods & Please enjoy our inaugural YouTube video about the creation of my album with Shooter Jennings, right here: https://t.co/N1RReBLopn & 4271926 \\
\hline
13 & TheBabylonBee & The Babylon Bee & \begin{tabular}[c]{@{}l@{}}Fake news you can trust.\\ January 6: The Most Deadliest Day— now streaming!\end{tabular} & 4217074 \\
\hline
14 & RobertKennedyJr & Robert F. Kennedy Jr &  & 4110129 \\
\hline
15 & PeteButtigieg & Pete Buttigieg & Personal account. For official updates, follow @SecretaryPete. Husband, father, veteran, writer, South Bend’s former Mayor Pete. (he/him) & 3876190 \\
\hline
16 & charliekirk11 & Charlie Kirk & Founder \& CEO: @TPUSA • @TPAction\_ • Host: The Charlie Kirk Show • Click the link below to subscribe & 3689713 \\
\hline
17 & libsoftiktok & Libs of TikTok & News you can’t see anywhere else. submissions@libsoftiktok.com. DM submissions. Bookings: Partnerships@libsoftiktok.com. Subscribe to our newsletter & 3619947 \\
\hline
18 & InternetH0F & internet hall of fame & the internet just wouldn't be the same without these iconic posts. & 3360230 \\
\hline
19 & megynkelly & Megyn Kelly & Happily married to Doug, crazy in love with my children Yates, Yardley, and Thatcher, journalist. & 3278907 \\
\hline
20 & catturd2 & Catturd ™ & The turd you can’t flush. & 3054117 \\
\hline
21 & ProjectLincoln & The Lincoln Project & ``You cannot escape the responsibility of tomorrow by evading it today.'' – Abraham Lincoln | Home of \#TheBreakdown and LP Podcast & 2994780 \\
\hline
22 & hodgetwins & Hodgetwins & Merch \& Giveaways at: https://t.co/kxb8qjGCDW  —— PODCAST: @thetwinspod & 2993141 \\
\hline
23 & bennyjohnson & Benny Johnson & i make internet & 2937906 \\
\hline
24 & JackPosobiec & Jack Posobiec & Sr Editor, @HumanEvents. Veteran Navy intel officer. Catholic. NYT Bestselling Author & 2809915 \\
\hline
25 & gtconway3d & George Conway & President and Executive Director of @PsychoPAC24, the Anti-Psychopath Political Action Committee. President, @chkbal. Contributor, @TheAtlantic. & 2396447 \\
\hline
26 & unusual\_whales & unusual\_whales & \begin{tabular}[c]{@{}l@{}}Stocks/Options/Crypto/Market News + Tools. Not advice \\ Get \$50-\$5000 to trade: https://t.co/wGf2ZdlXpw\\ Discord: https://t.co/0xJ9e0ZYYG\\ More: https://t.co/nsxZlPV0pC\end{tabular} & 1901961 \\
\hline
27 & PopCrave & Pop Crave & Craving Pop Culture. & 1884374 \\
\hline
28 & DC\_Draino & DC\_Draino & Rogan O’Handley & 1855070 \\
\hline
29 & DiscussingFilm & DiscussingFilm & Your leading source for quick reliable news. Home for healthy and liberating discussion on all things pop culture. (Amazon links shared may earn us commissions) & 1835626 \\
\hline
30 & ScottPresler & ThePersistence & I helped defeat Hillary, Cheney, \& organized the Baltimore cleanup. My goal is to re-elect President Trump. Check out @EarlyVoteAction  MAGA MAHA & 1776345 \\
\hline
31 & TheRickWilson & Rick Wilson & Lincoln Project. Award-winning ad-maker. Writer. Instrument-rated pilot. NYT \#1 best-seller. Still got the shovel. Writing: https://t.co/e04n749N5H & 1698059 \\
\hline
32 & PopBase & Pop Base & Pop Base is your best source for all pop culture related entertainment, news, award show coverage, chart updates, statistics and more. | email@popbase.tv & 1683990 \\
\hline
33 & CollinRugg & Collin Rugg & Co-Owner of Trending Politics | Investor | American & 1561596 \\
\hline
34 & atensnut & Juanita Broaddrick & Author, ``You'd Better Put Some Ice On That'' retired RN \& business owner, Speaker. & 1455919 \\
\hline
35 & KamalaHQ & Kamala HQ & Providing context. & 1416761 \\
\hline
36 & kylegriffin1 & Kyle Griffin & Executive Producer @TheWeekendMSNBC. Opinions mine. Do not congratulate. THREADS @griffinkyle & 1409244 \\
\hline
37 & joncoopertweets & Jon Cooper & Ex: LI Campaign Chair for Barack Obama; National Finance Chair of Draft Biden; Majority Leader of Suffolk County Legislature. Gay dad of 5 kids. \#YesWeKam & 1391095 \\
\hline
38 & LauraLoomer & Laura Loomer & Investigative Journalist Free Spirit Founder of LOOMERED. Host of @LoomerUnleashed Former @Project\_Veritas operative. America First Feisty Jewess & 1364113 \\
\hline
39 & Tim\_Walz & Tim Walz & Running to win this thing with @KamalaHarris. & 1311484 \\
\hline
40 & Riley\_Gaines\_ & Riley Gaines & Host of Gaines for Girls podcast | Author of Swimming Against the current | TPUSA contributor | Director of the Riley Gaines Center & 1283591 \\
\hline
41 & MeidasTouch & MeidasTouch & The official account of the MeidasTouch Network. Unapologetically pro-democracy. & 1239469 \\
\hline
42 & RexChapman & Rex Chapman & It’s Hard For Me to Live With Me is available now.  For speaking inquiries please contact Jornstein@wmeagency.com & 1221843 \\
\hline
43 & AdamKinzinger & Adam Kinzinger & Proud RINO, dad, Husband, Lt. Col in @AirNatlGuard, CNN Senior Political Commentator, former Congressman, founder @thecountryfirst & 1082499 \\
\hline
44 & Scaramucci & Anthony Scaramucci &  Entrepreneur @SkyBridge. Host, Open Book and @RestPoliticsUS. https://t.co/t4SOzQjxuy & 1077632 \\
\hline
45 & JoJoFromJerz & Jo & mom. jersey. dem. news junkie. Lebanese. hothead.views are my own.https://t.co/zueo7YDFWx https://t.co/q4qgmwRLzt. https://t.co/9Fp1kdOX6w & 1029714 \\
\hline
46 & RonFilipkowski & Ron Filipkowski & Editor-in Chief https://t.co/HLS0hEHY1C, Co-host Uncovered, Attorney, Marine, Former Federal and State Prosecutor, Republican Party Insane Asylum Escapee & 1021928 \\
\hline
47 & GuntherEagleman & Gunther Eagleman™ & Political Commentator - America First - MAGA - Trump 2024 - Unfiltered & 1011785 \\
\hline
48 & atrupar & Aaron Rupar & journalist. sign up for my newsletter, Public Notice (link below). Powered by @SnapStream (more info: https://t.co/2oHPuuFBnN). & 987623 \\
\hline
49 & Dexerto & Dexerto & The leading source for influencer, streamer, gaming, and viral content & 980351 \\
\hline
50 & cb\_doge & DogeDesigner & UX/UI \& Graphic Designer at Dogecoin \& MyDoge Inc./ Citizen Journalist & 935600 \\
\hline
51 & marceelias & Marc E. Elias & Founder @DemocracyDocket. Chair @EliasLawGroup. My dog's name is Bode. & 899982 \\
\hline
52 & RpsAgainstTrump & Republicans against Trump & Pro-democracy conservatives Republicans fighting Trump \& Trumpism. Please support our work: https://t.co/FkmisNic4X & 821564 \\
\hline
53 & MuellerSheWrote & Mueller, She Wrote & \begin{tabular}[c]{@{}l@{}}DONATE to Kamala Harris: https://t.co/gOvFmy1bYN\\ Subscribe to my FREE newsletter\end{tabular} & 803719 \\
\hline
54 & harris\_wins & Kamala’s Wins & Keeping Score of Kamala Harris’ wins. The largest online community supporting soon to be President Kamala Harris & 790310 \\
\hline
55 & LeadingReport & Leading Report & Leading source for breaking news. & 630544 \\
\hline
56 & Angry\_Staffer & Angry Staffer & Not a WH Staffer | Politics, NatSec, and Snark - Your Mileage May Vary | Subscribe to my Patreon newsletter for free: https://t.co/Kj4zTIcPyk | & 609103 \\
\hline
57 & TristanSnell & Tristan Snell & Lawyer, legal commentator, fighter for democracy. Prosecuted Trump University @ NY AG. Commentator, MSNBC. Creator of book/podcast/newsletter TAKING DOWN TRUMP. & 583266 \\
\hline
58 & 7Veritas4 & Jack E. Smith & “Whatever you are, be a good one”. Here for people, politics and PARODY. alt @jackesmith22 & 543575 \\
\hline
59 & Victorshi2020 & Victor Shi & Now—Working on Team Harris-Walz. Writer. Fmr—Host @iGenPolitics\_, @JoeBiden, @WhiteHouse, @Precisionstrat, @SKDK. @UCLA 24 English alum. Chicagoan. Views mine. & 328827 \\
\hline
60 & acnewsitics & Alex Cole & \begin{tabular}[c]{@{}l@{}}Software Engineer \& Pilot | Progressive\\ Follow @newsitics \& https://t.co/Retehye9rD\end{tabular} & 286475 \\
\hline
61 & Logically\_JC & John Collins & Dad Husband  Low-Key Nerd  EdD / JD & 225758 \\
\hline
62 & scottlincicome & Scott Lincicome & @CatoInstitute Vice President (Econ/Trade), @DukeLaw adjunct, @TheDispatch newsletter-er. CH RTS. You didn't read the article, did you? Go @Rangers. & 78894 \\
\hline
63 & EpochTimesChina & The Epoch Times - China Insider & \begin{tabular}[c]{@{}l@{}}China content of The Epoch Times.\\ Sign up for our China newsletter\\ Read on App: https://t.co/wGG3L4uBaT\end{tabular} & 63943 \\
\hline
64 & Kalshi & Kalshi & The first legal way to bet on the election in America. & 50930 \\
\hline
65 & GanJingWorld & Gan Jing World & Video and movie streaming. Join \#KindnessIsCool contest \& win awards. Connect with friends \& family. & 32626 \\
\hline
66 & canlesofficial & Canles &  Engineered for walking |  Comfy \& versatile footwear for life's adventures |  Breathable, lightweight designs & 7407 \\
\hline
67 & janicehisle & Janice Hisle Epoch Times & Assigned to report on President Trump's 2024 campaign and related topics. Supporter of free speech. Email tips to janice.hisle@epochtimes.us. & 2846


\end{longtable}

\end{document}